\RequirePackage{fix-cm}
\documentclass[twocolumn]{svjour3}          
\smartqed  
\usepackage{graphicx}
\usepackage{bm}
\usepackage{siunitx}
\usepackage{hyphenat}
\usepackage{amsmath}	
\usepackage{amssymb}
\DeclareMathOperator{\tr}{Tr}

\newcommand{\figref}[1]{\figurename\,\ref{#1}}

\newcommand{\phiSJ}{\phi_{\mathrm{SJ}}}

\newcommand{\phiIJ}{\phi_{\mathrm{IJ}}}

\usepackage[usenames,dvipsnames]{xcolor}	
\begin{document}
\title{Shear jamming and fragility in dense suspensions}

\author{Ryohei~Seto \and
Abhinendra~Singh \and Bulbul~Chakraborty
\and Morton~M.~Denn \and Jeffrey~F.~Morris
}


\institute{R. Seto \at
  Department of Chemical Engineering, Kyoto University \\
  \email{setoryohei@me.com}           
  \and
  A. Singh \at
  Benjamin Levich Institute, CUNY City College of New York\\
  \emph{Present address: Institute for Molecular Engineering 
and James Franck Institute, The University of Chicago}
  \and
  B. Chakraborty \at
  Martin Fisher School of Physics, Brandeis University
  \and
  M. M. Denn
  \and
  J. F. Morris \at
  Benjamin Levich Institute and Department of Chemical Engineering, CUNY City College of New York,
}

\date{}
\maketitle

\begin{abstract}

The phenomenon of shear-induced jamming is a factor in the complex rheological behavior of dense suspensions.
Such shear-jammed states are fragile,
i.e., 
they are not stable against applied stresses that are incompatible with the stress imposed to create them.
This peculiar flow-history dependence of the stress response
is due to flow-induced microstructures.
To examine jammed states realized under constant shear stress,
we perform dynamic simulations of 
non-Brownian particles
with frictional contact forces
and hydrodynamic lubrication forces.
We find clear signatures 
that distinguish these fragile states from the more conventional isotropic jammed states.

\end{abstract}

\section{Introduction}
\label{intro}

Suspensions, in which solid particles are dispersed in a viscous liquid,
are a class of complex fluids found
frequently in nature, industry, 
and consumer applications\,\cite{Guazzelli_2018,Denn_2018}.
To predict flows of suspensions with arbitrary macroscopic boundary conditions,
it is necessary to develop continuum models based on particle-scale physics;
it is too expensive to simulate individual motions 
of particles and interstitial flows for macroscopic problems.
Dilute suspensions, in which the solid volume fractions $\phi$ are less than about 5\%, 
are well described with the Newtonian constitutive model 
with a modified viscosity~\cite{Einstein_1906,Einstein_1911}.
However, 
constitutive models 
for denser suspensions exhibiting more complex rheological properties
are still not available~\cite{Denn_2014,Goddard_2006}.
%


Suspensions are always liquid-like fluids
below a certain volume fraction,
i.e., there is no possibility to realize states exhibiting rigidity by any protocols.
Conversely, it is possible to induce rigidity in suspensions 
above a certain solid volume fraction.
Shear jamming is the phenomenon
when shear strains yield such a rigidity~\cite{Bertrand_2002,Peters_2016,Singh_2018}.
\begin{figure}
\centering
\includegraphics[width=0.38\textwidth]{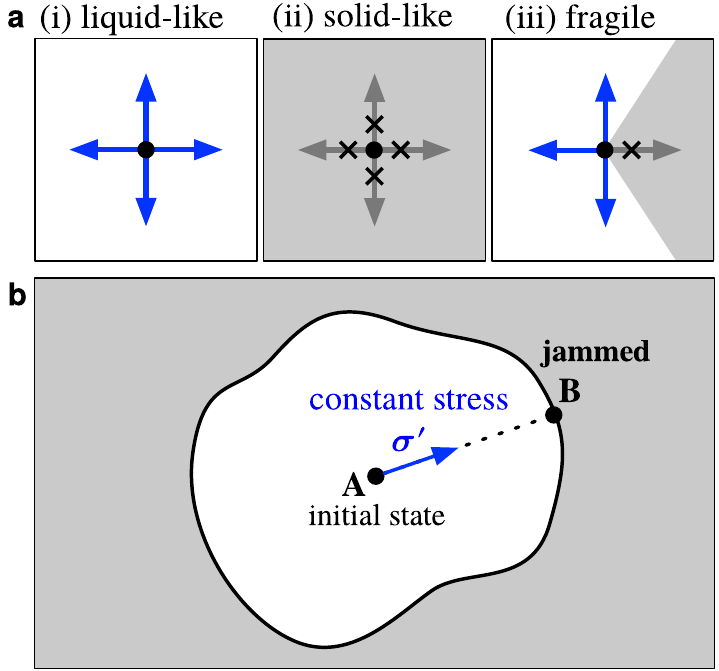} 
\caption{%
\textsf{\textbf{a}}~Three possible states of non-Brownian suspensions:
(i) liquid-like, (ii) solid-like, and (iii) fragile,
are defined according to distinct stress responses.
Different directions 
indicate stresses of different compression/traction axes
in this representation.
\textsf{\textbf{b}}~A schematic ``phase space'' of microstates (structures) of a suspension.
A suspension is initially in a liquid-like state,
represented by the point \textbf{A}.
It flows under constant stress $\bm{\sigma}'$ and reaches to a jammed state $\mathbf{B}$. 
The gray domain is unreachable with constant stresses from $\mathbf{A}$.
}
\label{fig_fragility_concept}
\end{figure}
\emph{Fragile matter} as a constitutive class of complex fluids
was introduced to 
describe emergence of rigidity in flowing suspensions~\cite{Cates_1998a,Cates_2000a}.
First, it is helpful to emphasize that there are only three possible states
in non-Brownian suspensions of rigid particles regarding mechanical responses 
(see \figref{fig_fragility_concept}\,\textsf{\textbf{a}}):
\begin{enumerate}
 \item[(i)] 
\emph{liquid-like} state, that cannot statically bear any shear stress.

\item[(ii)] \emph{solid-like} state, that can statically bear stresses in all directions.

\item[(iii)] \emph{fragile} state, that can statically bear stresses 
only within a certain range of directions.
\end{enumerate}
We assume sufficiently weak stresses (or infinitely rigid particles)
in these classifications to exclude yielding.
The unjammed states are liquid-like,
and isotropically jammed states are solid-like.
A number of processes may lead to these two states.
On the other hand, fragile states are usually associated with 
particular processes leading to fragile configurations.
Let us consider an idealized system of rigid-particle suspensions.
If we randomly pick a relaxed state (i.e., without flow-induced microstructure)
 below the isotropic jamming point,
it should be liquid-like and flow under arbitrary shear stress $\bm{\sigma}'$.
In the schematic configuration space shown 
in \figref{fig_fragility_concept}\,\textsf{\textbf{b}},
such an initial state is expressed as a point $\mathbf{A}$.  
The trajectory (dashed line) indicates the shearing process due to $\bm{\sigma}'$, 
which passes through different particle configurations.
The flow induces some microstructure to resist the applied stress,
which slows down the flow eventually bringing it to a stop;
i.e., the system reaches a jammed state, $\mathbf{B}$.
This jammed state statically supports the stress $\bm{\sigma}'$, 
like an elastic or rigid solid; unlike deformed elastic materials,
however, 
it is able to maintain the strain 
even after the stress is no longer applied.
This jammed state is unstable against a change in the applied stress.
Since we reached the jammed state $\mathbf{B}$ with the stress $\bm{\sigma}'$,
we may reverse the deformation
with the opposite stress $-\bm{\sigma}'$, at least to some extent.
Thus, jammed states encountered in shearing processes 
under constant stresses seem to always be fragile.
(In general, such jammed states may be able to support stress reoriented 
at a small angle of shear direction~\cite{Ness_2018} 
or stress with the principal axes rotated in a small angle. 
However, the existence of one incompatible stress is enough to 
judge fragility; thus, we consider only shear stress reversal here.)
If a different stress were to be applied to the initial state $\mathbf{A}$
from the beginning,
the system would reach another jammed state.
Such jammed states from $\mathbf{A}$
with different applied stresses form a surface,
beyond which configurations are unreachable from $\mathbf{A}$.
%


A shear-jammed state is one which is 
reached by shear but is then able to statically bear the shear stress (or `load').
As this state is not statically stable for applied stresses incompatible with the jammed state, 
it is termed fragile in the terminology of 
Cates \textit{et al.} \cite{Cates_1998a}, and we follow that terminology here; 
\figref{fig_fragility_concept}\,\textsf{\textbf{a}} (iii) illustrates the concept.  Upon reversal of the shear stress from a fragile shear-jammed state, the suspension 
will flow, i.e., undergo some finite strain, before possibly reaching a jammed state in the new direction.


The occurrence of shear jamming under quasi-statically imposed strain 
was experimentally elucidated in frictional grains by Bi \textit{et al.}~\cite{Bi_2011},
and its connection to Reynolds dilatancy was elaborated in Ren \textit{et al.}~\cite{Ren_2013}.
Two types of shear-jammed states were identified at a given $\phi$:
states created at lower strains, which could not sustain shear reversal, 
and states at strains higher than a characteristic value, which could.
In~\cite{Bi_2011}, the former were referred to as fragile and the latter as shear-jammed.
Sarkar \textit{et al.}~\cite{Sarkar_2013,Sarkar_2015,Sarkar_2016} developed 
a theoretical framework to describe the transition between 
the two types of jammed states identified in~\cite{Bi_2011}.
Recent numerical work by Otsuki and Hayakawa~\cite{Otsuki_2018} showed 
that this transition could be detected through imposition of oscillatory shear.
It should be noted that those previous studies on shear jamming were performed 
with strain-controlled protocols,
in which deformation is forced regardless of whether jammed or not.
In contrast, we investigate shear jamming with a stress-controlled protocol; 
once a system is jammed, no further deformation occurs in the same direction. 
This stress-controlled approach seems, in this way,
more natural to capture shear jamming than the previous works. 
(Very recently, Srivastava \textit{et al.}~\cite{Srivastava_2019}
also investigated ``shear-arrested states''
using constant-stress discrete element simulations.)
In this article, we examine fragility of shear jammed states 
by performing particle dynamics simulations with idealized conditions: 
inertialess, non-Brownian, and monolayer systems. 

\section{Simulation model}

We consider suspended particles in a viscous liquid.
The particles are sufficiently small for all inertial effects to be negligible,
i.e., the Stokes number $\mathit{St} \equiv \rho a^2 \dot{\gamma}/\eta_0$
(with $\rho$ and $a$ being the density and radius of particles, respectively,
$\dot{\gamma}$ being the shear rate, and $\eta_0$ being the viscosity of the suspending fluid)
is vanishingly small.
We also omit Brownian motions,
which are relevant for smaller particles.
Stokesian Dynamics (SD) is an efficient method
to reproduce particle dynamics in this Stokes regime~\cite{Brady_1988}.
The viscosity divergence predicted by the original SD 
is a dynamic effect due to 
the singularity of hydrodynamic lubrication~\cite{Melrose_1995,Morris_2018}.
Recently, the SD approach (with only hydrodynamic lubrication)
was extended to be coupled with frictional contact mechanics
to reproduce discontinuous shear thickening \cite{Seto_2013a}.
In this work, we employ an algorithm to mimic stress-controlled rheology~\cite{Mari_2015}.
The viscosity divergence under a constant shear stress
is just a consequence of a static force balance of the contact forces.
In contrast to the original SD,
the hydrodynamic contributions vanish at the viscosity divergence.
Therefore, the results shown in this article
would share some common features with dry granular systems in the quasi-static limit.

\paragraph{Stress-controlled quasi-static dynamics}

Particles with negligible inertia suspended in a viscous liquid 
obey the force and torque balance equations of hydrodynamic 
and non-hydrodynamic interactions,
\begin{equation}
 \bm{F}_{\mathrm{h}}(\bm{U}) + \bm{F}_{\mathrm{nh}} = \bm{0},\label{094131_23Nov18}
\end{equation}
where $\bm{U}$ is the many-body linear and angular velocities of particles. 
In the zero-Reynolds number limit,
the hydrodynamic interactions can be expressed 
as a linear resistance,
$ \bm{F}_{\mathrm{h}}(\bm{U})  = -\boldsymbol{\mathsf{R}}_{\mathrm{FU}} \cdot \bm{U}$,
where $\boldsymbol{\mathsf{R}}_{\mathrm{FU}}$ is the resistance matrix~\cite{Brady_1988}.
Thus, particles are moved with $\bm{U} = \boldsymbol{\mathsf{R}}_{\mathrm{FU}}^{-1} \bm{F}_{\mathrm{nh}}$.
For the case of very dense suspensions,
$\boldsymbol{\mathsf{R}}_{\mathrm{FU}}$ can be approximately constructed
with the pair-wise hydrodynamic lubrication~\cite{Ball_1997,Jeffrey_1984}.
The lubrication coefficients are known to diverge when two spherical particles approach,
but we regularize these interactions with a cutoff length~\cite{Wilson_2002}.


With a background flow gradient of $\nabla \bm{u}$,
the hydrodynamic interactions
are modified to the sum of the linear resistances to 
the particle velocity deviations $\bm{U}-\bm{u}$ and to the rate-of-deformation tensor 
$\boldsymbol{\mathsf{D}} \equiv (\nabla \bm{u}+\nabla \bm{u}^{\mathsf{T}})/2$,
\begin{equation}
  \bm{F}_{\mathrm{h}} (\bm{U})
  = - \boldsymbol{\mathsf{R}}_{\mathrm{FU}} \cdot (\bm{U}-\bm{u}) 
  + \boldsymbol{\mathsf{R}}_{\mathrm{FD}} : \boldsymbol{\mathsf{D}},\label{094135_23Nov18}
\end{equation} 
where $\boldsymbol{\mathsf{R}}_{\mathrm{FD}}$ is also a resistance matrix~\cite{Mari_2014}.
Furthermore,
the simulation cell with periodic boundary conditions
needs to be deformed according 
to $\nabla \bm{u}$ (for more details see~\cite{Seto_2017}).


As a consequence of the linearity of the governing equations,
particle velocities $\bm{U}$ and the flow rate of a fixed flow type 
can be simultaneously determined under a given shear stress $\sigma^{xy}$.
Here, we fix the flow type to simple shear flows, 
$\bm{u}(\bm{r})  = \dot{\gamma} y \bm{e}_x$,
with shear rate $\dot{\gamma}$,
which is the only degree of freedom to be determined in $ \nabla \bm{u}$.
The stress tensor can be expressed as
the sum of the deformation contribution
and contributions by non-hydrodynamic interactions,
\begin{equation}
  \bm{\sigma} =
  \dot{\gamma} \hat{\bm{\sigma}}_{\mathrm{D}}
  + \bm{\sigma}_{\mathrm{nh}},
  \label{100951_23Nov18}
\end{equation}
in which the unknown shear rate $\dot{\gamma}$ is explicitly 
factored out from the first term.
The rest is independent of $\dot{\gamma}$,
\begin{equation}
 \hat{\bm{\sigma}}_{\mathrm{D}}
=  \frac{1}{V} \sum_i
\bigl(\boldsymbol{\mathsf{R}}_{\mathrm{SD}} :\hat{\boldsymbol{\mathsf{D}}}
+ \boldsymbol{\mathsf{R}}_{\mathrm{SU}} \cdot \hat{\bm{U}}_{\mathrm{D}}\bigr)^{(i)}, 
\end{equation}
with $\hat{\bm{U}}_{\mathrm{D}} = \boldsymbol{\mathsf{R}}_{\mathrm{FU}}^{-1} 
\cdot \boldsymbol{\mathsf{R}}_{\mathrm{FD}}:\hat{\boldsymbol{\mathsf{D}}}$
and the normalized rate-of-deformation tensor $\hat{\boldsymbol{\mathsf{D}}}
 \equiv \boldsymbol{\mathsf{D}} / \dot{\gamma} $.
The non-hydrodynamic contribution,
\begin{equation}
  \bm{\sigma}_{\mathrm{nh}}
  =  \frac{1}{V}
\biggl\{
\sum_{i>j} (\bm{r}_i - \bm{r}_j) \bm{F}^{(ij)}_{\mathrm{nh}}
- \sum_i (\boldsymbol{\mathsf{R}}_{\mathrm{SU}}\cdot\bm{U}_{\mathrm{nh}})^{(i)}
\biggr\},\label{180102_6Dec18}
\end{equation}
is also independent of $\dot{\gamma}$,
where 
$\bm{U}_{\mathrm{nh}} = \boldsymbol{\mathsf{R}}_{\mathrm{FU}}^{-1} \cdot \bm{F}_{\mathrm{nh}} $.
From the $xy$ component of \eqref{100951_23Nov18},
we can determine $\dot{\gamma}$ for the given shear stress $\sigma^{xy}$,
\begin{equation}
\dot{\gamma}
= \frac{\sigma^{xy} -\sigma_{\mathrm{nh}}^{xy} }{
\hat{\sigma}_{\mathrm{D}}^{xy}}.\label{135141_23Nov18}
\end{equation}
Now, we can also determine the particle velocities with the obtained $\dot{\gamma}$:
$\bm{U} = \dot{\gamma}\hat{\bm{U}}_{\mathrm{D}} + \bm{U}_{\mathrm{nh}}$.

\paragraph{Contact force model}

In many stable suspensions used in practice, 
some repulsive forces act between non-contacting particles,
preventing flocculation due to short-range van der Waals attractions.
However, in this article,
we focus on a simple model system 
in which the non-hydrodynamic interaction consists only of contact forces:
$\bm{F}_{\mathrm{nh}} = \bm{F}_{\mathrm{c}}$.
%


To model $\bm{F}_{\mathrm{c}}$,
we employ a soft-constraint approach.
The first geometrical constraint 
is the volume excluding force of solid particles.
Hard-sphere particles will have zero overlap.
To mimic this,
we introduce a harmonic penalty function
$(k_{\mathrm{n}}/2)(a_i+a_j-r_{ij})^2$,
which generates a force along the normal direction.
Here, $a_i$ and $a_j$ are radii of particles $i$ and $j$,
and $r_{ij}$ the distance between them.
%


When rough or bumpy solid surfaces are in contact,
their sliding displacements are also restricted by friction or interlocking; 
here, we introduce another harmonic penalty function $(k_{\mathrm{t}}/2)\bm{\xi}^2$
of the relative sliding displacement $\bm{\xi}$,
which is determined with translations and rotations of contacting particles~\cite{Luding_2008}.
This generates tangential forces acting at the contact point.
Regarding the maximum tangential force, 
we employ a simple Coulomb friction law, 
where the upper bound is proportional to the normal force 
with a proportionality coefficient $\mu$. 
In this work, we mainly study an infinite friction coefficient,
implying that sliding displacements are constrained
as long as the particles are pushed inward.


%
This soft-constraint approach is fundamentally different from hard-sphere algorithms, 
which impose strict geometrical constraints.
To make the constraints stricter,
we need to set sufficiently large values
for the penalty parameters $k_{\mathrm{n}}$ and $k_{\mathrm{t}}$.
The values which we selected keep the maximum overlap and tangential displacement
less than 2\% of the particle radius below the isotropic jamming point.

\section{Results and discussion}

We study monolayer bidisperse systems of 1000 spherical particles, 
with a size ratio of $1.4$ and a volume ratio of approximately $1$. 
To generate initial configurations,
we used Brownian simulations 
to relax randomly placed particle configurations.

\paragraph{Shear reversal test}
\begin{figure}
\centering
\includegraphics[width=0.4\textwidth]{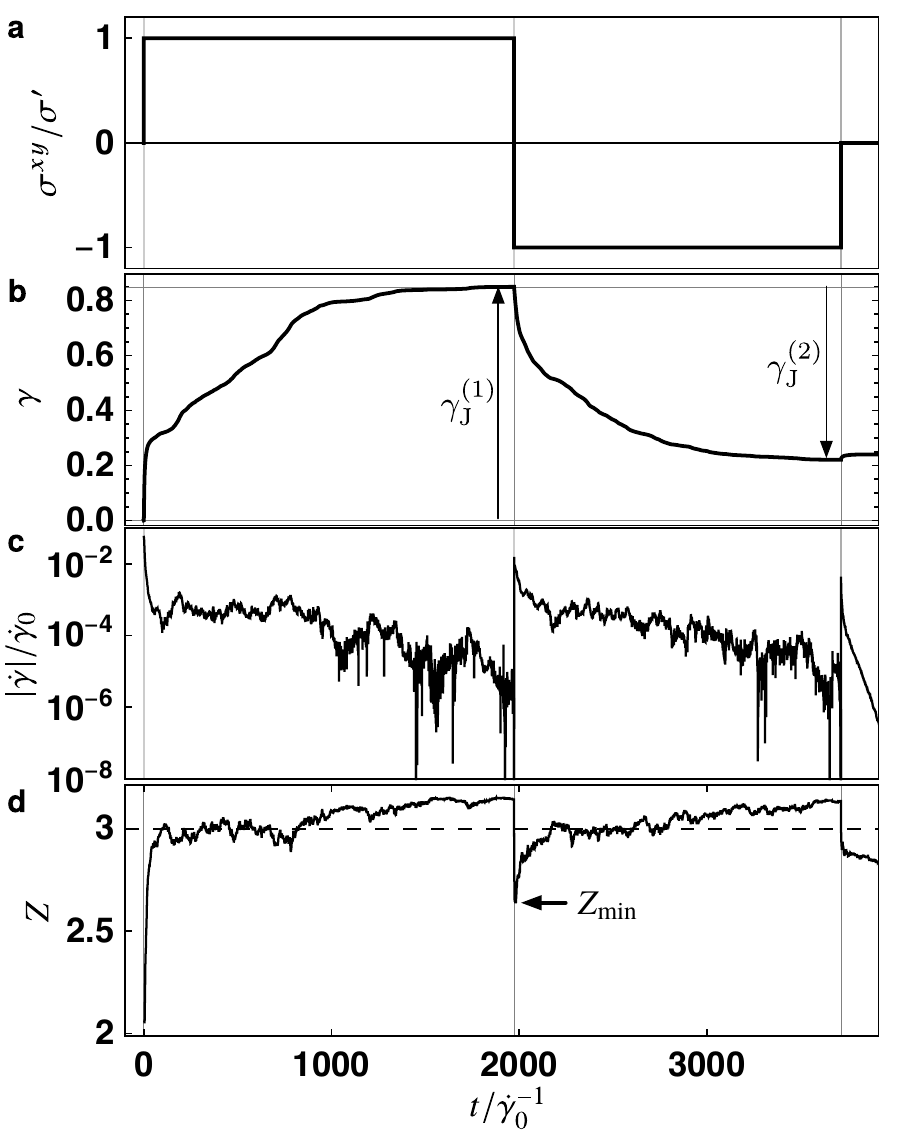} 
\caption{%
Time series of (\textsf{\textbf{a}}) shear stress $\sigma^{xy}$,
(\textsf{\textbf{b}}) shear strain $\gamma$,
(\textsf{\textbf{c}}) absolute value $|\dot{\gamma}|$ of shear rate, 
and (\textsf{\textbf{d}}) contact number $Z$ with non-rattlers 
in a stress-controlled shear reversal test at $\phi = 0.77$ with $\mu = \infty$.
The shear stress is reversed after reaching the jammed state.
After reaching the second jammed state, the shear stress is stopped.
The unit of shear rate $\dot{\gamma}_0 \equiv \sigma'/ \eta_0$
is used for the nondimensionalization.
}
\label{fig_shear_reversal}
\end{figure}

We start by confirming the concept of fragile matter with 
our simulation model for dense suspensions.
To understand the roles of shear-induced structure,
Gadala-Maria and Acrivos~\cite{GadalaMaria_1980}
performed shear reversal tests using a rate-controlled setup.
Here, we simulate a stress-controlled shear reversal test.
%


We apply a constant stress $\sigma^{xy} = \sigma'$ 
to an equilibrated suspension of $\phi = 0.77$\,(\figref{fig_shear_reversal}\,\textsf{\textbf{a}}).
The strain evolution is relatively fast at the beginning 
of the simulation and eventually slows down\,(\figref{fig_shear_reversal}\,\textsf{\textbf{b}}).
The slowly flowing state, say $|\dot{\gamma}| > 10^{-4}\dot{\gamma}_0$,
lasts for a while.
Fluctuation of $\dot{\gamma}$ in the flowing state
indicates some restructuring of the stress-bearing 
contact network~(\figref{fig_shear_reversal}\,\textsf{\textbf{c}}).
%

The system is shear-jammed when all particles are in static 
force balance $\bm{F}_{\mathrm{C}}^{(i)} = 0$
and a contact network to support all stress is formed 
such that $\sigma^{xy}_{\mathrm{C}} = \sigma^{xy}$.
According to \eqref{094131_23Nov18} and \eqref{135141_23Nov18},
these conditions lead to $\bm{U}^{(i)} = 0$ and $\dot{\gamma} = 0$.
We consider states to be jammed with the following criteria:
$\max |\bm{V }^{(i)}| < 10^{-3}a \dot{\gamma}_0$ and $|\dot{\gamma}| < 10^{-8} \dot{\gamma}_0$; the characteristic shear rate $\dot{\gamma}_0 \equiv \sigma'/ \eta_0$
is used.
Here, $\bm{V}^{(i)} \equiv \bm{U}^{(i)}-\bm{u}(\bm{r}^{(i)})$ are 
non-affine particle velocities.
$\gamma_{\mathrm{J}}^{(1)}$ denotes the total strain to the shear-jammed\,(SJ) states
from the relaxed initial configuration.
Jamming occurs above the isostatic condition 
$Z > Z_{\mathrm{iso}}^{\mu=\infty} = 3$ \cite{Henkes_2010},
where $Z$ is the average contact number with non-rattlers.
Particles that have fewer than two contacts with non-rattlers
are called rattlers,
and thus we need some iteration to determine them.
As seen in \figref{fig_shear_reversal}\,\textsf{\textbf{d}},
the isostatic condition does not immediately lead to the SJ state.


Now, we reverse the shear stress to $\sigma^{xy}= -\sigma'$, corresponding to 
a rotation of the principal stress axes by $\pi/2$.
Since the previously formed contact network cannot support this new stress, the suspension unjams.
The contact number drops to a value $Z_{\mathrm{min}}$, being below the isostatic condition (\figref{fig_shear_reversal}\,\textsf{\textbf{d}}).
Thus, the stress-reversal simulation confirms that the SJ state is \emph{fragile}.
We continue the simulation with $-\sigma'$.
The particle dynamics is not reversible,
and the state does not return to the initial configuration; rather, 
it reaches another SJ state after strain $\gamma_{\mathrm{J}}^{(2)}$.


After reaching the second jammed state,
we stop applying the stress $\sigma^{xy} = 0 $ to confirm 
the smallness of the elastic recovery strain (\figref{fig_shear_reversal}\,\textsf{\textbf{b}}).
This small recovery is due to 
the finite values of the penalty parameters 
$k_{\mathrm{n}}$ and $k_{\mathrm{t}}$ in the soft-constraint contact model.
In the ideal hard-sphere limit, the recovery strain will be zero.
If stress is applied in the same direction again,
the system will not flow because the contact network remains.

\begin{figure*}
\centering
  \includegraphics[width=0.88\textwidth]{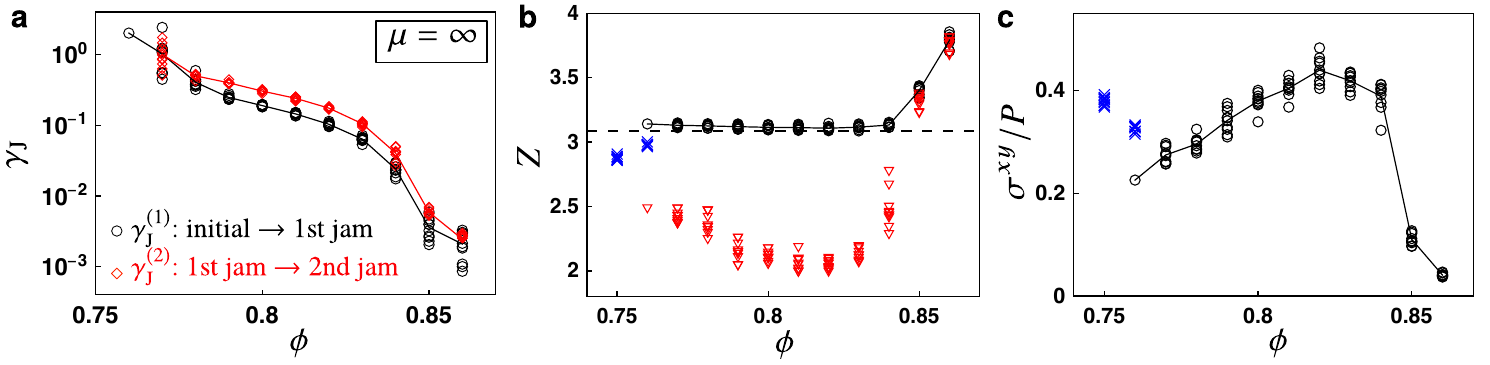}
\caption{%
\textsf{\textbf{a}}~The larger is $\phi$, the smaller is the average strain $\gamma_{\mathrm{J}}$ 
to reach a SJ state.
$\gamma_{\mathrm{J}}^{(1)}$ ($\circ$)
and $\gamma_{\mathrm{J}}^{(2)}$ (\textcolor{red}{$\diamond $})
are strains to reach the first jammed states from the initial states
and the second jammed states after stress reversals, respectively.
Only jammed results of ten simulations are plotted.
%
%
\textsf{\textbf{b}}~Mean contact number $Z$ with non-rattlers
of SJ states (\textcolor{black}{$\circ$})
are almost constant for $\phi \leq 0.84$.
The lowest value of the SJ states is $ Z \approx 3.07$ (dashed line).
These SJ states can be confirmed as fragile
with the minimum values $Z_{\mathrm{min}}$ 
after the shear reversals (\textcolor{red}{$\triangledown$})
(see \figref{fig_shear_reversal} \textsf{\textbf{d}}),
which are below the isostatic condition $Z_{\mathrm{iso}} = 3$
for $\phi \leq 0.84$.
$Z$ of unjammed states (\textcolor{blue}{$\times$})
are below but close to the plateau value near the boundary.
%
%
\textsf{\textbf{c}}~The sharp decrease of the stress anisotropy $\sigma^{xy}/P$
of jammed states (\textcolor{black}{$\circ$})
above $\phi = 0.84$ indicates the transition from shear jamming to isotropic jamming~\cite{Sarkar_2013,Sarkar_2015,Sarkar_2016}.
}
\label{fig_FJ_data}       
\end{figure*}

\paragraph{Features of shear-jammed states}

The SJ state is realized
due to formation of a stress-bearing contact network.
The particle movements obey the force and torque balance equations~\eqref{094131_23Nov18},
and rearrangements continue until static balances are globally achieved.
The structural evolution to reach the global balance is not monotonic.
As seen in the movie in Supplementary Material,
static force balance,
which is roughly indicated by vanishing velocities (dark colors),
is locally achieved in advance of other parts,
but the local stress axes may change due to rearrangements of other parts.
Thus, the local domains, which achieved force balance once, 
need to be restructured again (cf. ``micro-fragility'' in \cite{Cates_2000a}).
%


The strain $\gamma_{\mathrm{J}}$ to reach a SJ state reflects
the difficulty in realizing the global force balance.
\figref{fig_FJ_data}\,\textsf{\textbf{a}} shows the area fraction dependence
of $\gamma_{\mathrm{J}}^{(1)}$ from relaxed initial configurations 
to the first jammed states,
and $\gamma_{\mathrm{J}}^{(2)}$ 
from the first to the second jammed states.
Particle contacts to build a network are more accessible at higher area fractions,
and shear jamming accordingly occurs at smaller strain.
All simulations for $\phi \geq 0.77$ indeed end up in jammed states,
but require larger $\gamma_{\mathrm{J}}$ for lower $\phi$.
Only one of ten simulations at $\phi = 0.76$, 
and none at $\phi = 0.75$, were jammed within the given maximum strain $\gamma_{\mathrm{max}} = 5$.
Thus, the threshold area fraction $\phiSJ$ to realize SJ states
is expected to be in the range $ 0.75 < \phiSJ < 0.76$, 
although based on the results available we cannot rule 
out the possibility of eventual shear jamming at $\phi = 0.75$ or even lower.
%


Though the isostatic condition $Z = Z_{\mathrm{iso}}$ ($= 3 $ for $\mu = \infty$) 
alone does not determine whether or not shear jamming occurs,
shear jamming was realized at slightly larger $Z \approx 3.1$
in all of our simulations for $\phi \leq 0.84$ 
(\figref{fig_FJ_data}\,\textsf{\textbf{b}}).
If we were to run more simulations with larger $\gamma_{\mathrm{max}}$, 
the minimum line of $Z$ might approach the isostatic condition.
(The SJ state at the lowest possible $\phi$ may be close 
to random loose packing~\cite{Onoda_1990,Ciamarra_2008}, 
but is anisotropic owing to the shearing by which it is accessed.)
We can confirm fragility 
with the minimum value (\textcolor{red}{$\triangledown$}) 
of $Z$ after the stress reversal.
If $Z$ goes below $Z_{\mathrm{iso}}$,
the system must experience liquid-like states 
no matter how short their duration.

\paragraph{Isotropic jamming}

Strains to achieve jammed states
become progressively lower at higher area fractions (\figref{fig_FJ_data}\,\textsf{\textbf{a}}).
The strains $\gamma_{\mathrm{J}}^{(1)}$ are less than 0.01 at $\phi \geq 0.85$,
and higher penalty parameters for the contact model, $k_{\mathrm{n}}$ and $k_{\mathrm{t}}$,
can make them even lower (data are not shown).
The vanishing value of $\gamma_{\mathrm{J}}^{(2)}$ suggests
that the state does not flow in any direction,
i.e., the jammed state is solid-like.
As seen in \figref{fig_FJ_data}\,\textsf{\textbf{c}},
the stresses of these states indeed become more isotropic 
(The ratio $\sigma^{xy} / P $
is one way to represent the stress anisotropy,
where $P$ is the particle pressure~\cite{Giusteri_2018}).
We can also see the sudden increase in $Z$ above $\phi = 0.85$~(\figref{fig_FJ_data}\,\textsf{\textbf{b}}). 
These observations suggest that the solid-like jammed states are 
obviously distinguishable from the SJ state, 
but similar to the conventional isotropically jammed state
despite the presence of friction.
The transition point $\phiIJ$ seems to be in the range $ 0.84 < \phiIJ < 0.85$,
which agrees with the known value (about $0.84$)
for frictionless 2D systems~\cite{Behringer_2019}.


\paragraph{Stress-bearing structure}

\begin{figure}
  \centering
  \includegraphics[width=0.48\textwidth]{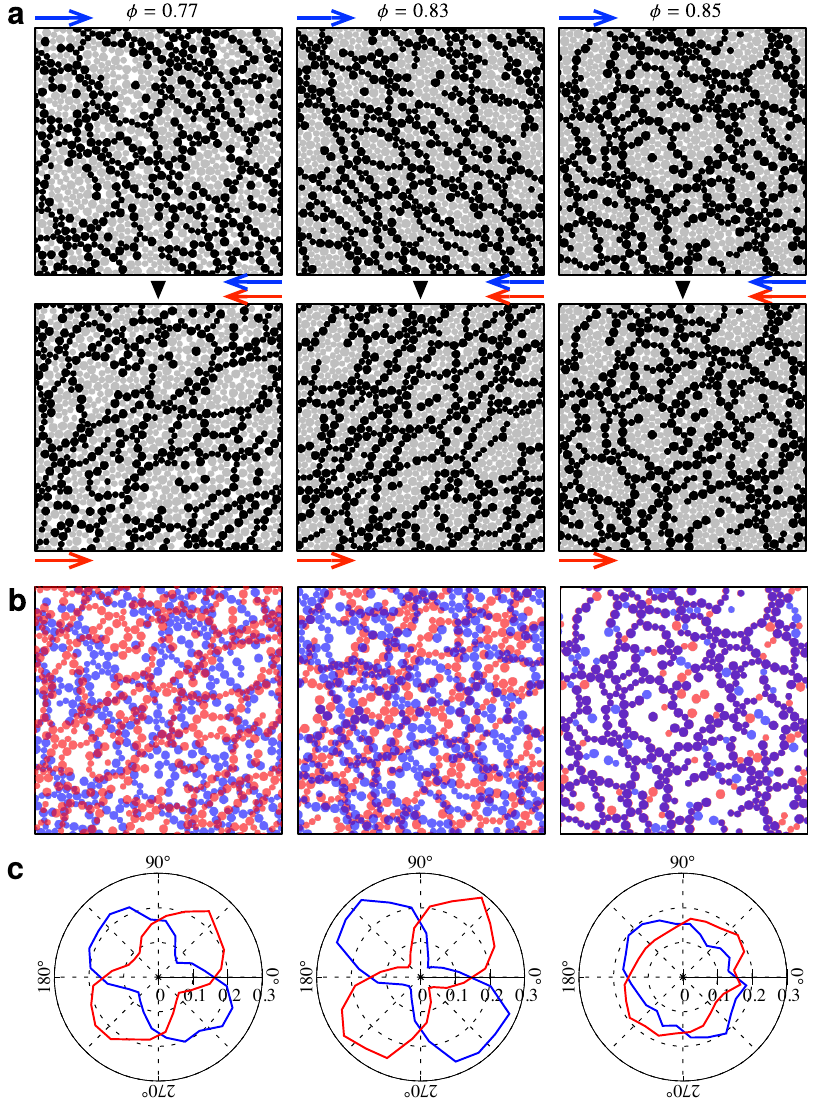}
  \caption{%
    \textsf{\textbf{a}}
    Stress transmission patterns of the first jammed states under $\sigma^{xy}= \sigma'$ 
    (upper) and the second jammed states under $\sigma^{xy}= -\sigma'$ (lower) are shown.
    Stressed particles, $P^{(i)} > \langle P \rangle$, are in black, 
    where $P^{(i)} \equiv -\tr \bm{\sigma}^{(i)}/2$ is particle pressure of the $i$-th particle.
    \textsf{\textbf{b}}
    Stressed particles in the first (blue) and second (red) 
    jammed states are imposed.
    Overlapping particles appear in purple.
    \textsf{\textbf{c}}
    The polar plots show the probabilities
    of the orientation between two contacting stressed-particles 
    in the first (blue) and the second (red) jammed states.
    The distributions are obtained from 10 simulations.
    The polar plots 
    grow more anisotropic from $\phi = 0.77$ to $0.83$, 
    but become more isotropic at $\phi = 0.85$.
  }
\label{fig_network}       
\end{figure}

\figref{fig_network}\,\textsf{\textbf{a}} shows stress transmission patterns 
bearing $\sigma'$ in the first jammed states (upper)
and $-\sigma'$ in the second jammed states (lower), respectively.
The superposition of stressed particles 
in the first and second jammed states 
(\figref{fig_network}\,\textsf{\textbf{b}})
displays how stress bearing structures
are different at $\phi = 0.77$ and $0.83$ 
but more similar at $\phi = 0.85$.
The anisotropy, 
which may be noticed at $\phi = 0.77$ and $0.83$ in \figref{fig_network}\,\textsf{\textbf{a}},
is confirmed with the angular distributions 
for the orientation of contacting stressed particles (\figref{fig_network}\,\textsf{\textbf{c}}).
In this way,
fragile SJ states require some compatible anisotropic structures;
thus they are renewed to adapt to the opposite direction of the applied stress.

\paragraph{Friction}

\begin{figure}
  \centering
  \includegraphics[width=0.47\textwidth]{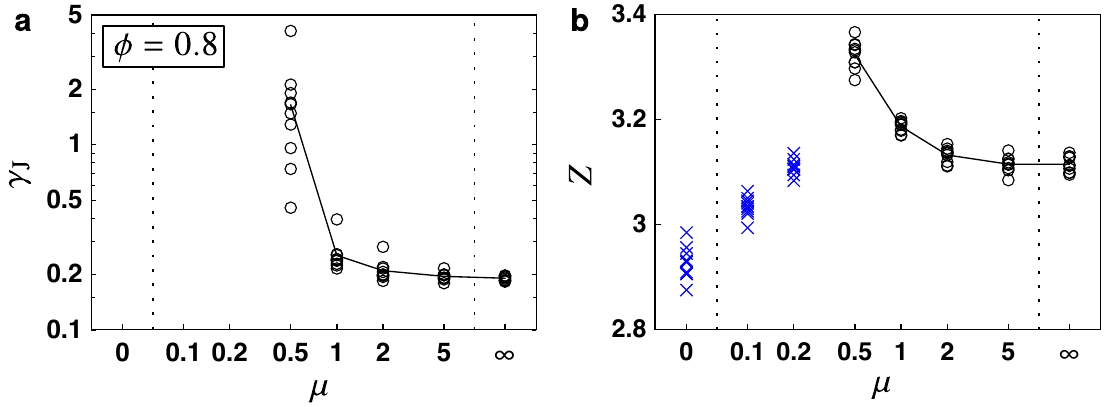}
  \caption{
    \textsf{\textbf{a}}~Friction coefficient $\mu$ dependence
    of the shear jamming strain $\gamma_{\mathrm{J}}$ 
    (\textcolor{black}{$\circ$}) at $\phi = 0.8$.
    Suspensions with lower frictions ($\mu = 0.2$ and below) did not reach jamming
    at this area fraction.
    \textsf{\textbf{b}}~The contact number $Z$ (\textcolor{blue}{$\times$}) of 
    unjammed states increases with the friction coefficient $\mu$.
    However, 
    lower $Z$ (\textcolor{black}{$\circ$}) is required to realize jammed states
    with higher $\mu$.
  }
  \label{fig_fric_dependence}
\end{figure}

So far we have focused on the theoretical limit of frictional systems with $\mu = \infty$.
We briefly discuss the $\mu$ dependence of our results.
As seen in \figref{fig_fric_dependence}\,\textsf{\textbf{a}}, 
the shear jamming was achieved only when $\mu \geq 0.5$ at $\phi = 0.8$,
which indicates that $\phiSJ$ shifts to higher values with 
a weaker friction $\mu$, as expected.
The strain $\gamma_{\mathrm{J}}$ to reach SJ states
increases with smaller $\mu$.
More contacts $Z$ are required 
to realize jamming (\figref{fig_fric_dependence}\,\textsf{\textbf{b}}).
We also notice that
the contact number $Z$ of unjammed states increases with $\mu$;
particles tend to contact in frustrated flows due to friction.

\begin{figure}
  \centering
  \includegraphics[width=0.47\textwidth]{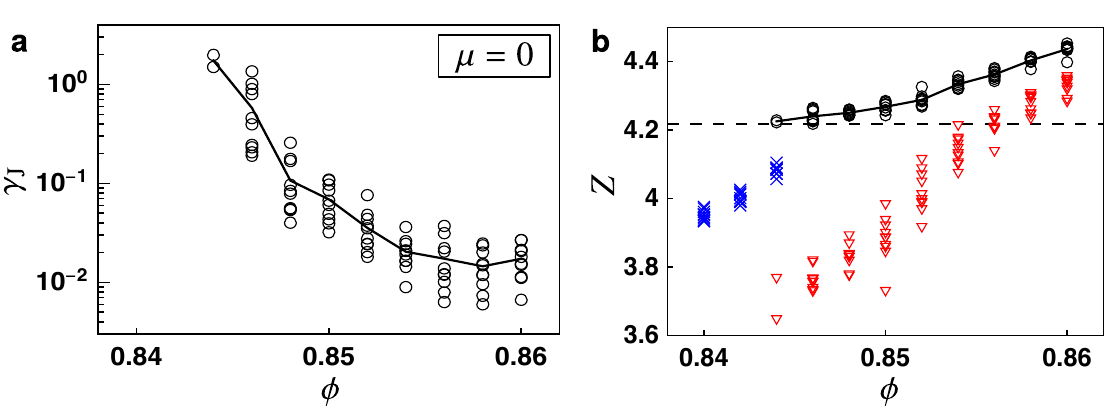}
  \caption{\textsf{\textbf{a}} Strains $\gamma_{\mathrm{J}}$
    to reach jammed states for frictionless suspensions ($\mu = 0$).
    \textsf{\textbf{b}} The average contact numbers $Z$ (\textcolor{black}{$\circ$})
    monotonically increase as the volume fraction $ \phi $.
    SJ states may be indicated
    by the minimum values after the shear reversal (\textcolor{red}{$\triangledown$})
    which are below the isostatic condition $Z_{\mathrm{iso}}=4$.
    However, the observed range of area fractions is rather narrow.
    $Z$ of unjammed states are time-averaged values (\textcolor{blue}{$\times$}).
  }
\label{fig_frictionless}
\end{figure}

Even when friction is completely absent ($\mu = 0$),
we obtained a similar shear-jamming phenomenology;
the systems are shear-jammed after some flow (\figref{fig_frictionless}\,\textsf{\textbf{a}}),
and the contact numbers $Z$ of jammed states 
drop to below the isostatic condition ($Z_{\mathrm{iso}}^{\mu=0} = 2d$)
after the 
stress reversal (\figref{fig_frictionless}\,\textsf{\textbf{b}}).
However, this occurs in a narrow range just below the isotropic jamming.
Our particles seem too soft to see a clear transition 
from the shear jamming to the isotropic jamming.
As discussed elsewhere~\cite{Baity-Jesi_2017},
the shear jamming observed in frictionless particles can be just due to a finite size effect.
Thus, our current simulation cannot confirm the existence 
of shear jamming without friction.

\section{Conclusions}

We confirmed that
dense suspensions with frictional interactions between particles can behave as fragile matter.
In a flowing dense suspension under stress,
a contact network is formed.
The suspension becomes jammed when the shear-induced 
contact network statically supports the entire stress.
However, this jammed state is not stable;
a change of stress axes makes it flow.
This fragile instability is the most important feature 
of shear jamming
to be distinguished from the solid-like isotropic jamming.
Furthermore, we found various signatures
to distinguish the two different states
in the average contact number $Z$,
the drop of $Z$ after the stress reversal,
and the stress anisotropy $\sigma^{xy}/P$.
It is also worth noting that SJ states near the lower bound 
are truly ``fragile.''
We need to set a sufficiently short time-step
to capture such SJ states 
in simulations with the soft-constraint contact model.
%


In this article,
we did not investigate the dependence 
on the strength $|\sigma^{xy}|$ of the shear stress.
Ideal inertialess hard-sphere suspensions
do not possess any internal force scale;
thus, the states must be independent of the stress scale.
Therefore,
in the phase diagram with stress and area fraction,
shear jamming lies
in the vertical boundaries: $\phiSJ < \phi < \phiIJ$.
If some interparticle repulsive forces or Brownian forces act on particles,
they tend to hinder the formation of a contact network.
This competition introduces a stress dependence.
Deformability of particles also causes a similar stress dependence;
contact deformation can enhance tangential constraints
restricting sliding and rolling degrees of freedom~\cite{Cates_1998a}.
It is worth noting the distinction
between shear thickening and Reynolds dilatancy here~\cite{Barnes_1989}.
Shear thickening does require such an internal force scale
besides tangential constraints;
this makes the rheology of suspensions rate dependent~\cite{Seto_2013a}.
Shear jamming is relevant to shear thickening 
but is a more basic phenomenon;
it can occur just due to shear strain
without the internal force scale, 
as demonstrated in this article.
Since the volume of a suspension is constrained, 
shear jamming of dense suspensions 
can be considered as a confined Reynolds dilatancy~\cite{Reynolds_1885}.

\begin{acknowledgements}
The authors would like to thank M. Otsuki, H.\,Hayakawa, and R.\,Mari for fruitful discussions.
This study was supported by the Japan Society for the Promotion of Science 
(JSPS) KAKENHI Grants No.\,17H01083 and No.\,17K05618.
BC was supported by NSF-CBET-1605428, while JFM was supported by NSF-CBET-1605283.
The research was also supported in part by the National Science Foundation 
under grant No.\,NSF PHY-1748958. RS thanks R.\,Yamamoto for his full support.
\end{acknowledgements}

\bibliographystyle{spphys}       

\end{document}